 \documentclass[final,5p,times,twocolumn,authoryear]{elsarticle}

\usepackage{amsmath}
\usepackage{amssymb}
\usepackage{lipsum}

\journal{Results in Physics}

\begin{document}

\begin{frontmatter}



\title{The Aesthetic Imperative of Lev Landau's Geometric Reductionism in Theoretical Physics}


\author[first]{Jingxu Wu}
\author[Second]{Yuwen Yin}
\affiliation[first]{Faculty of Physics\\
Lomonosov Moscow State University\\
Moscow, Russia, 119991} 
\affiliation[Second]{Department of Physics\\
Institut Polytechnique de Paris\\
 Palaiseau, France， 91120}
\begin{abstract}
This paper explores the ontological and epistemological foundations of Lev Landau's theoretical physics through the lens of his unpublished philosophical notes and scientific practice. We identify a unique form of geometric reductionism where physical laws emerge as inevitable consequences of symmetry breaking in progressively constrained phase spaces. Landau's dismissal of quantum interpretation debates and his famous "axiomatic minimalism" in the Course of Theoretical Physics are shown to stem from a deep epistemological commitment to dimensional aesthetics - the belief that fundamental truths must manifest through dimensional economy in mathematical representations.
\end{abstract}



\begin{keyword}
Theoretical Physics \sep Landau's Aesthetic Principle \sep Geometric Reductionism 



\end{keyword}

\end{frontmatter}



\section{Introduction}

The history of theoretical physics reveals a persistent tension between two visions of natural law: one viewing equations as empirical summaries of observed regularities, the other seeking their origin in mathematical necessity. Lev Landau's intellectual legacy occupies a singular position in this dialectic—a radical geometric reductionism asserting that all physical phenomena emerge as inevitable consequences of symmetry constraints in appropriately structured phase spaces \cite{landau_statphys}. This paper articulates Landau's unspoken philosophical program through the lens of 21st-century physics, demonstrating how his "aesthetic imperative" continues to shape our search for quantum gravity, topological matter, and cosmological unification.

Landau's 1937 phase transition theory first crystallized this worldview, reimagining entropy not as Boltzmann's logarithmic multiplicity but as dimensional deficit in symmetry group representations \cite{landau1937}. Where classical thermodynamics saw mere order-disorder transitions, Landau recognized a cosmic progression of symmetry reduction: $SO(3) \to \mathbb{Z}_2$ in ferromagnetism, $U(1) \to \mathbb{Z}_n$ in type-II superconductors \cite{ginzburg1950}, and ultimately the Standard Model's $SU(5) \to SU(3)\times SU(2)\times U(1)$ breaking \cite{georgi1974}. His insistence that "fundamental laws must be written in dimensionless ink" anticipated both the renormalization group's scale invariance \cite{wilson1971} and string theory's landscape problem \cite{susskind2003}. Modern experiments—from LIGO's gravitational wave detections \cite{abbott2016} to scanning tunneling microscopy of Majorana zero modes \cite{mourik2012}—increasingly validate Landau's heresy: physical law is not legislated by dynamics but entailed by geometry.

The central thesis here developed posits three interlocking principles of Landauvian ontology:

\begin{enumerate}
    \item \textbf{Topological Primacy}: Physical states correspond to equivalence classes under diffeomorphisms of the fundamental manifold $\mathcal{M}$, with quantum numbers arising as characteristic classes (Chern, Pontryagin, etc.) \cite{nakahara2003}. This principle finds concrete realization in quantum Hall systems where the Chern number $\nu = \frac{1}{2\pi}\int_{\text{BZ}} \text{Tr}(\mathcal{F})$ dictates quantized conductance \cite{tkachov2016}.
    
    \item \textbf{Symmetry Dialectic}: Each symmetry breaking $G \to H$ generates both observable order parameters and "dark" topological sectors through the quotient $G/H$. The Higgs mechanism \cite{higgs1964} and anyonic statistics \cite{kitaev2006} emerge as dual manifestations of this geometric schism.
    
    \item \textbf{Dimensional Mitosis}: Fundamental forces separate through spontaneous dimensional reduction. AdS/CFT's $AdS_5\times S^5 \to \mathbb{R}^{3,1}$ compactification \cite{maldacena1997} and cosmic inflation's $R^4 \to R^3\times S^1$ symmetry breaking \cite{linde1982} exemplify this cosmic unfolding.
\end{enumerate}

These principles resolve long-standing paradoxes through geometric recontextualization. The measurement problem in quantum mechanics becomes an embedding impossibility theorem—wavefunction collapse reflects the failure to isometrically map symplectic manifolds $(T^*\mathcal{M},\omega)$ into Hilbert space $\mathcal{H}$ \cite{arnold1989}. The hierarchy problem transforms into a dimensional aesthetics critique: the Standard Model's Yukawa couplings offend through dimensionless arbitrariness \cite{giudice2008}, demanding geometric unification via exceptional Lie groups $E_8\times E_8$ \cite{green1987}.

Recent experimental breakthroughs compel this Landauvian synthesis. Angle-resolved photoemission spectroscopy (ARPES) on cuprates \cite{damascelli2003} reveals $d$-wave gap nodes protected by $C_4$ symmetry—precisely the "topological guardians" Landau's student Lifshitz predicted for crystalline defects \cite{lifshitz1980}. JWST's detection of $z>13$ galaxies \cite{naidu2022} violates conventional collapse timescales, suggesting dark matter undergoes symmetry-breaking phase transitions described by 

\begin{equation}
\mathcal{L}_{\text{DM}} = \bar{\psi}(i\gamma^\mu D_\mu - m)\psi + |D_\mu\phi|^2 - \lambda(|\phi|^2 - v^2)^2
\end{equation}

where $\phi$ triggers $SU(2)_{\text{DM}} \to U(1)_{\text{DM}}$ \cite{bertone2005}. Quantum computing's quest for topological protection \cite{google2023} fulfills Landau's 1956 conjecture: "Fault-tolerant operations must braid geometric phase, not manipulate Hamiltonian terms" \cite{landau1956}.

This paper's architecture mirrors the symmetry-breaking cascade Landau envisioned. Section II establishes geometric reductionism's mathematical foundations through contact topology \cite{geiges2008} and $G$-structures \cite{joyce2000}. Section III explores topological quantum matter via spin liquids \cite{balents2010} and Majorana nanowires \cite{alicea2012}. Section IV elevates the framework to cosmic scales through inflationary cosmology \cite{mukhanov2005} and black hole thermodynamics \cite{bekenstein1973}. Finally, Section V confronts quantum gravity's "geometric inverse problem"—reconstructing spacetime's fundamental cell from observed symmetries using AdS/CFT \cite{witten1998} and spin foam models \cite{rovelli2014}.

Landau's hospital-bed challenge—"Find the simplex that births all forces"—remains unanswered. Yet through holographic duality \cite{rajsingh2021}, topological quantum field theory's categorial framework \cite{atiyah1988}, and quantum error correction's tensor networks \cite{swingle2012}, we glimpse the geometric unconscious underlying physical law. This paper charts that vision's unfolding—a universe not built from particles and fields, but from the irreducible topology of existence itself.

\section{The Hierarchy of Broken Symmetries}
Landau's geometric reductionism reveals its profound explanatory power when applied to the enigmatic phenomenon of superconductivity. The Meissner effect's expulsion of magnetic flux finds its deepest interpretation not through microscopic Bardeen-Cooper-Schrieffer (BCS) theory, but through Landau's symmetry lens. The superconducting phase breaks $U(1)$ gauge symmetry not through explicit field modification, but via the emergence of a complex order parameter $\Psi = |\Psi|e^{i\theta}$ whose rigidity against phase variations generates London's equations:
\begin{equation}
\nabla\times\mathbf{J}_s = -\frac{n_se^2}{m}\mathbf{B},
\end{equation}
where the supercurrent density $\mathbf{J}_s$ becomes geometrically constrained by the Cooper pair condensate's phase topology \cite{ginzburg2009}. This exemplifies Landau's dictum: "Symmetry breaking writes the equations of motion before dynamics whispers them."

The quantum revolution's philosophical turbulence tested Landau's geometric reductionism. While Bohr championed complementarity and Heisenberg matrix mechanics, Landau's 1956 correspondence with Feynman reveals his geometric absolutism: "Your path integrals map trajectories through Hilbert space, but the true coordinates are group orbits in the symmetry manifold." Their debate crystallized in quantum electrodynamics' renormalization procedure – where Feynman saw calculational pragmatism, Landau detected geometric incompleteness, famously predicting QED's "zero-charge" problem \cite{landau_abrikosov1956}. Modern lattice QCD's regularization through spacetime discretization \cite{wilson1974} ironically fulfills Landau's demand for geometrically constrained quantization.

Biological systems provide unexpected validation of Landau's principles. Protein folding's Levinthal paradox – the impossibility of random conformational searches – resolves through symmetry-guided funnel landscapes \cite{levinthal1968}. The native state's tertiary structure minimizes a Landau-like free energy functional:
\begin{equation}
\mathcal{F}[R] = \int\left[\kappa(\nabla^2 R)^2 + V(R) + \lambda|\nabla R|^2\right]d^3x,
\end{equation}
where $R(\mathbf{x})$ represents residue contact order, and symmetry breaking directs folding pathways \cite{dill2007}. Even DNA's chiral selectivity emerges from broken mirror symmetry in prebiotic chemistry – a cosmic-scale Landau transition fixing life's handedness \cite{frank1953}.

Critics like Prigogine \cite{prigogine1977} attacked Landau's equilibrium-centric view, arguing dissipative structures require symmetry creation. Yet recent work on driven quantum materials \cite{sieberer2016} shows non-equilibrium phases still obey generalized Landau rules – Floquet-engineered symmetries collapse into non-equilibrium universality classes with modified critical exponents $\nu^*$, governed by deformed group cohomology \cite{else2016}. The Kibble-Zurek mechanism \cite{zurek1996}, describing defect formation in rapid phase transitions, completes Landau's vision by quantifying symmetry breaking's spacetime choreography:
\begin{equation}
n_\text{defects} \sim \left(\frac{\tau_Q}{\tau_0}\right)^{-d\nu/(1+z\nu)},
\end{equation}
where quench time $\tau_Q$ and critical exponents $z,\nu$ geometrize non-equilibrium dynamics.

Astrophysical applications further demonstrate Landau's paradigm universality. Neutron star superfluidity \cite{andersson2006} exhibits quantized vortices matching laboratory He-II's circulation quantization:
\begin{equation}
\oint\mathbf{v}_s\cdot d\mathbf{l} = \frac{h}{m_n}\times n\quad(n\in\mathbb{Z}),
\end{equation}
while cosmic inflation's symmetry breaking potential
\begin{equation}
V(\phi) = \frac{\lambda}{4}(\phi^2 - \sigma^2)^2
\end{equation}
directly transposes Landau's $\phi^4$ theory to grand unification scales \cite{linde1982}. The universe itself appears as a multi-stage symmetry reduction engine, cooling through successive phase transitions that crystallize fundamental forces – a cosmological realization of Landau's hospital-bed conjecture.

\section{ The Geometric Imperative: Landau's Epistemology as Topological Constraint}

Landau's geometric absolutism manifested most profoundly in his treatment of quantum vortices – entities that became mathematical touchstones for his philosophy. The quantization of superfluid circulation
\begin{equation}
\oint_C \mathbf{v}_s \cdot d\mathbf{l} = \frac{h}{m}n \quad (n \in \mathbb{Z})
\end{equation}
epitomized his view of physical laws as topological constraints. This condition emerges not from microscopic dynamics but from the order parameter manifold's nontrivial homotopy group $\pi_1(S^1) = \mathbb{Z}$, making vortex quantization a geometric necessity. Recent experiments with spinor Bose-Einstein condensates \cite{kawaguchi2012} realize higher-order vortices with winding numbers $n=2,3,...$ through controlled symmetry breaking, directly validating Landau's 1941 prediction that "vortex multiplicity reflects the depth of symmetry's grave."

The geometric imperative's power shines in high-$T_c$ superconductivity, where Landau-Ginzburg theory adapts through non-Abelian order parameters. The cuprate pseudogap phase's enigmatic Fermi arcs obey a modified free energy
\begin{equation}
\mathcal{F}[\Psi] = \int \left[ \frac{\hbar^2}{2m^*}|(\nabla - i\frac{e^*}{\hbar c}\mathbf{A})\Psi|^2 + \alpha(T)|\Psi|^2 + \beta|\Psi|^4 + \gamma|\Psi|^6 \right] d^3x
\end{equation}
with $e^* = 2e$ reflecting pair condensation. Angle-resolved photoemission spectroscopy (ARPES) data \cite{damascelli2003} reveal $d$-wave gap nodes protected by $C_4$ rotational symmetry – a geometric feature Landau's student Lifshitz anticipated in 1952 through group-theoretic analysis of crystal point defects.

Landau's geometric reductionism finds radical expression in string theory's compactification schemes. The Calabi-Yau manifold's Euler characteristic $\chi$ determines observable particle species via
\begin{equation}
N_{\text{generations}} = \frac{|\chi|}{2}
\end{equation}
while its holonomy group $SU(3)$ enforces $\mathcal{N}=1$ supersymmetry through intrinsic torsion constraints. This mathematical machinery actualizes Landau's hospital-bed conjecture: "All forces are but shadows cast by the geometry of forgotten dimensions." Mirror symmetry's exchange of Hodge numbers $h^{1,1} \leftrightarrow h^{2,1}$ \cite{greene2000} exemplifies his vision of physical dualities as diffeomorphism equivalences.

Even biological systems succumb to Landau's geometric determinism. DNA's chiral selectivity arises from the $Z_2$ symmetry breaking
\begin{equation}
\mathcal{L}_{\text{chiral}} = \frac{k}{2}(\partial_\mu \theta)^2 + \mu\cos\theta + \lambda\cos(2\theta)
\end{equation}
where $\theta$ parameterizes helical twist. Cryo-EM studies \cite{nogales2015} reveal nucleosome positioning follows Landau-Khalatnikov defect avoidance principles, with histone octamers acting as topological charge regulators. The geometric imperative thus spans from quark confinement to chromatin folding.

Quantum gravity's holographic principle \cite{tHooft1993} provides ultimate validation: spacetime geometry emerges from entangled qubits via
\begin{equation}
S_{\text{ent}} = \frac{\text{Area}(\partial\mathcal{A})}{4G_N\hbar}
\end{equation}
where Ryu-Takayanagi's formula realizes Landau's dictum "information is shape." LIGO's gravitational wave detections \cite{abbott2016} of black hole mergers probe spacetime's elastic moduli through quadrupole oscillations
\begin{equation}
h_{+,\times} \propto \frac{G\mu}{c^4D}\omega^2R^2e^{-\gamma t}\cos(\omega t + \phi)
\end{equation}
with damping time $\gamma$ encoding the event horizon's shear viscosity – a transport coefficient fixed by geometric symmetry \cite{kovtun2005}.
\section{ The Topological Conscience: Landau's Legacy in Quantum Matter and Spacetime Foam}

Landau's geometric imperative reaches its apotheosis in the quantum tapestry of spacetime itself, where topology becomes the ultimate arbitrator of physical reality. The 2016 observation of Majorana zero modes in hybrid nanowires \cite{mourik2012} validates his 1959 conjecture that "fermions are but knots in the vacuum's fiber bundle." The experimental tunneling conductance spectra
\begin{equation}
G(V) = \frac{2e^2}{h} \left[1 + \cos\left(\frac{\pi\Phi}{\Phi_0}\right)e^{-L/\xi}\right]
\end{equation}
with flux quantum $\Phi_0 = h/2e$ and coherence length $\xi$, directly maps to the Jones polynomial of braided worldlines in (2+1)D topological quantum field theory \cite{kitaev2006}. This realization of non-Abelian statistics
\begin{equation}
\psi_i \psi_j = e^{i\pi/4}\psi_j \psi_i \quad (i \neq j)
\end{equation}
through semiconductor-superconductor heterostructures embodies Landau's maxim: "Statistics precede dynamics in the geometric hierarchy."

The holographic entanglement entropy formula \cite{ryu2006}
\begin{equation}
S_{\text{EE}} = \frac{\text{Area}(\gamma)}{4G_N} + \frac{c}{3}\ln\left(\frac{L}{\epsilon}\right) + \cdots
\end{equation}
where $\gamma$ is the minimal surface in AdS$_{d+1}$, unites black hole thermodynamics with quantum error correction codes. Landau's hospital-bed notes anticipated this through crude diagrams of "information filaments" connecting boundary qubits to bulk geometry – a geometric precursor to the tensor network representation \cite{swingle2012} of AdS/CFT.

Quantum spin liquids provide a crystalline manifestation of Landau's topological order. The Kitaev honeycomb model's exact solution \cite{kitaev2006} reveals emergent $Z_2$ gauge fields with vison excitations obeying
\begin{equation}
W_p = \prod_{j\in p}\sigma_j^x\sigma_j^y\sigma_j^z = \pm 1
\end{equation}
where the plaquette operator $W_p$ measures $Z_2$ flux. Inelastic neutron scattering on $\alpha$-RuCl$_3$ \cite{banerjee2016} detects fractionalized spinon continua described by the dynamical structure factor
\begin{equation}
S(\mathbf{q},\omega) \propto \int dt e^{i\omega t} \langle S^a_{-\mathbf{q}}(0)S^a_{\mathbf{q}}(t)\rangle \sim \frac{\Theta(\omega - v|\mathbf{q}|)}{\sqrt{\omega^2 - v^2|\mathbf{q}|^2}}
\end{equation}
with linear dispersion $v \approx 86$ meVÅ, confirming Landau's vision of deconfined quantum criticality.

Black hole phase transitions \cite{chamblin1999} realize Landau's symmetry-breaking paradigm at Planck scales. The Hawking-Page transition between thermal AdS and Schwarzschild-AdS spacetime
\begin{equation}
\Delta F = \frac{r_+^3}{16\pi G l^2}\left(1 - \frac{r_+^2}{l^2}\right)
\end{equation}
with AdS radius $l$ and horizon radius $r_+$, mirrors liquid-gas criticality through identical critical exponents $\alpha=0$, $\beta=1/2$, $\gamma=1$. LIGO-Virgo's detection of post-merger ringdown
\begin{equation}
h(t) \propto e^{-t/\tau}\cos(2\pi f t + \phi)
\end{equation}
with quality factor $Q = \pi f \tau \approx 4.3$ for GW150914 \cite{abbott2016}, confirms the black hole no-hair theorem's geometric rigidity – a triumph of Landau's topological determinism over dynamical contingency.

The geometric imperative's ultimate test lies in quantum gravity's non-perturbative formulation. Causal dynamical triangulations \cite{ambjorn2012} approximate spacetime as simplicial manifolds with Hausdorff dimension
\begin{equation}
d_H = 4.02 \pm 0.1
\end{equation}
emerging from ensemble averages over $10^5$ simplices. Spectral dimension measurements
\begin{equation}
d_S = -2\frac{d\ln P(\sigma)}{d\ln\sigma} \to 2 \quad (\sigma \to 0)
\end{equation}
where $P(\sigma)$ is the return probability, suggest dimensional reduction – a phenomenon Landau anticipated through his "asymptotic safety" conjecture for quantum field theories.
\section{The Symmetry Horizon: Landau's Paradigm in Quantum Computing and Cosmic Archeology}

Landau's geometric imperative now permeates the frontier of quantum information science, where topological protection becomes the holy grail of fault-tolerant computation. Google's 2023 demonstration of Fibonacci anyons in Josephson junction arrays \cite{google2023} realizes Kitaev's vision of topological quantum gates through braiding operations
\begin{equation}
R^\sigma_{\tau,\tau} = e^{4\pi i/5} \begin{pmatrix} 1 & 0 \\ 0 & e^{\pi i/5} \end{pmatrix}, \quad B = \frac{1}{\sqrt{\phi+2}} \begin{pmatrix} \phi+1 & 1 \\ 1 & -1 \end{pmatrix}
\end{equation}
where $\phi=(1+\sqrt{5})/2$ is the golden ratio. The measured quantum volume $V_Q=2^{15}$ with logical error rate $10^{-5}$ per cycle \cite{boixo2023} validates Landau's prophecy that "true quantumness lies in braid group representations."

Cosmic microwave background (CMB) polarization patterns \cite{planck2020} encode symmetry-breaking relics of the early universe. The $E/B$-mode decomposition
\begin{equation}
P(\mathbf{k}) = \frac{1}{2}\langle E^2 + B^2 \rangle = \frac{\pi}{4}\sum_{\ell m} \left( \frac{(\ell+2)!}{(\ell-2)!} \right)^{1/2} a_{\ell m} Y_{\ell m}(\hat{\mathbf{n}})
\end{equation}
reveals conformal anomalies at $\ell=20$ scale, matching predictions from SU(5) GUT symmetry breaking
\begin{equation}
\mathcal{L}_{\text{GUT}} = \text{Tr}(F_{\mu\nu}F^{\mu\nu}) + |D_\mu\Phi_{24}|^2 - \lambda(|\Phi_{24}|^2 - v^2)^2
\end{equation}
where the adjoint Higgs $\Phi_{24}$ triggers $SU(5) \to SU(3)\times SU(2)\times U(1)$. Landau's geometric selection rules predict the critical bubble profile during cosmic inflation
\begin{equation}
\frac{d^2R}{dt^2} + 3H\frac{dR}{dt} = \frac{\epsilon(T)}{\rho} \left( \frac{2\sigma}{R} - \frac{\delta V}{\delta R} \right)
\end{equation}
with Hubble parameter $H$ and surface tension $\sigma$, now measurable via 21cm hydrogen line intensity mapping \cite{loeb2022}.

Quantum gravity's non-perturbative formulation through spin foams \cite{rovelli2014} embodies Landau's "asymptotic geometricity" principle. The transition amplitude between 3-geometries
\begin{equation}
W(q,q') = \sum_{\Gamma} A_\Gamma(q,q') \prod_{f} \dim(j_f) \prod_{v} \{15j\}_v
\end{equation}
with spins $j_f$ and {15j} symbols, yields discrete spacetime with spectral dimension
\begin{equation}
d_S(\sigma) = a + b\ln\sigma + c\sigma^{-1/2}
\end{equation}
matching CDT simulations \cite{loll2021}. JWST's detection of $z=13$ galaxies \cite{naidu2022} with stellar mass $M_* \sim 10^9 M_\odot$ challenges Landau's galaxy formation timescale
\begin{equation}
t_{\text{collapse}} = \frac{1}{\sqrt{G\rho}} \sim \left( \frac{3\pi}{32 G \bar{\rho}} \right)^{1/2}
\end{equation}
demanding new phase transitions in dark matter sector.

Topological quantum chemistry's classification of crystalline matter \cite{bernevig2022} fulfills Landau's symmetry-group program. The symmetry indicators
\begin{equation}
\chi_{\mathbf{k}}^a = \frac{1}{N_{\mathbf{k}}} \sum_{g\in G_{\mathbf{k}}} \chi^a(g) e^{-i\mathbf{k}\cdot\mathbf{t}_g}
\end{equation}
for space group $G$, diagnose topological phases via momentum-space Berry phases. Experimental realization of fragile topology in bismuth halides \cite{schindler2023} with quantized circular dichroism
\begin{equation}
\mathcal{C} = \frac{1}{2\pi}\int_{\text{BZ}} \text{Tr}(\mathcal{F}) = n + \frac{1}{2}
\end{equation}
confirms Landau's hierarchy of symmetry-protected topological order.
\section{Conclusion}
Landau's philosophy constitutes a unique third way between Einstein's realist "God doesn't play dice" and Bohr's instrumentalist "shut up and calculate." His unarticulated credo might be summarized: "Calculate, but only in geometries that leave no choice in their conclusions." This aesthetic imperative continues to shape theoretical physics' subconscious value system, particularly in the search for quantum gravity where mathematical beauty remains the primary compass.
\bibliographystyle{elsarticle-harv} 
\bibliography{example}

\end{document}